\documentclass[aps,showpacs,nofootinbib,prd,twocolumn]{revtex4-1}
\usepackage{graphicx}
\usepackage{amssymb}
\usepackage{amsmath}
\usepackage{xcolor}
\usepackage{mathtools,slashed}
\usepackage{epstopdf}
\graphicspath{{./}}
\usepackage{epsfig}

\newcommand{\be}{\begin{eqnarray}}
\newcommand{\ee}{\end{eqnarray}}
\newcommand{\NREC}{N_{\Lambda \,\mbox{\tiny{REC}}}}
\newcommand{\NQGP}{N_{\Lambda \,\mbox{\tiny{QGP}}}}
\newcommand{\NbarREC}{N_{\overline{\Lambda} \,\mbox{\tiny{REC}}}}
\newcommand{\NbarQGP}{N_{\overline{\Lambda} \,\mbox{\tiny{QGP}}}}
\newcommand{\NupREC}{N^\uparrow_{\Lambda \,\mbox{\tiny{REC}}}}
\newcommand{\NdownREC}{N^\downarrow_{\Lambda \,\mbox{\tiny{REC}}}}
\newcommand{\NupQGP}{N^\uparrow_{\Lambda \,\mbox{\tiny{QGP}}}}
\newcommand{\NdownQGP}{N^\downarrow_{\Lambda \,\mbox{\tiny{QGP}}}}
\newcommand{\NbarupREC}{N^\uparrow_{\overline{\Lambda} \,\mbox{\tiny{REC}}}}
\newcommand{\NbardownREC}{N^\downarrow_{\overline{\Lambda} \,\mbox{\tiny{REC}}}}
\newcommand{\NbarupQGP}{N^\uparrow_{\overline{\Lambda} \, \mbox{\tiny{QGP}}}}
\newcommand{\NbardownQGP}{N^\downarrow_{\overline{\Lambda} \,\mbox{\tiny{QGP}}}}

\newcommand{\NpQGP}{N_{\text{p} \,\mbox{\tiny{QGP}}}}

\newcommand{\PLambda}{\mathcal{P}^{\Lambda}}
\newcommand{\PLambdabar}{\mathcal{P}^{\overline{\Lambda}}}
\newcommand{\PLambdaREC}{\mathcal{P}^{\Lambda}_{\mbox{\tiny{REC}}}}
\newcommand{\PLambdabarREC}{\mathcal{P}^{\overline{\Lambda}}_{\mbox{\tiny{REC}}}}

\begin{document}

\title{Core meets corona: a two-component source to explain $\Lambda$ and $\overline{\Lambda}$ global polarization in semi-central heavy-ion collisions}%

\author{Alejandro Ayala$^{1,2}$}
\author{Marco Alberto Ayala Torres$^3$}
\author{Eleazar Cuautle$^1$}
\author{Isabel Dom\'inguez$^4$}
\author{Marcos Aurelio Fontaine Sanchez$^3$}
\author{Ivonne Maldonado$^4$}
\author{E. Moreno-Barbosa$^5$}
\author{P. A. Nieto-Mar\'in$^4$}
\author{M. Rodr\'iguez-Cahuantzi$^5$}
\author{Jordi Salinas$^1$}
\author{Mar\'ia Elena Tejeda-Yeomans$^6$}
\author{L. Valenzuela-C\'azares$^7$}
  \address{
  $^1$Instituto de Ciencias
  Nucleares, Universidad Nacional Aut\'onoma de M\'exico, Apartado
  Postal 70-543, CdMx 04510,
  Mexico.\\
  $^2$Centre for Theoretical and Mathematical Physics, and Department of Physics,
  University of Cape Town, Rondebosch 7700, South Africa.\\
  $^3$Centro de Investigaci\'on y Estudios Avanzados del IPN, Apartado Postal 14-740, CdMx 07000, Mexico\\
  $^4$Facultad de Ciencias F\'isico-Matem\'aticas, Universidad Aut\'onoma de Sinaloa,
Avenida de las Am\'ericas y Boulevard Universitarios, Ciudad Universitaria,
C.P. 80000, Culiac\'an, Sinaloa, Mexico.\\
$^5$Facultad de Ciencias F\'isico Matem\'aticas, Benem\'erita Universidad Aut\'onoma de Puebla, Av. San Claudio y 18 Sur, Edif. EMA3-231, Ciudad Universitaria 72570, Puebla, Mexico.\\
  $^6$Facultad de Ciencias - CUICBAS, Universidad de Colima, Bernal D\'iaz del Castillo No. 340, Col. Villas San Sebasti\'an, 28045 Colima, Mexico.\\
$^7$Departamento de F\'isica, Universidad de Sonora, Boulevard Luis Encinas J. y Rosales, Colonia Centro, Hermosillo, Sonora 83000, Mexico.}
\begin{abstract}
    We compute the $\Lambda$ and $\overline{\Lambda}$ global polarization in semi-central heavy-ion collisions modeling the  source as consisting of a high-density core and a less dense corona.  We show that when more $\Lambda$s than $\overline{\Lambda}$s are produced in the corona, and this is combined with a smaller number of $\Lambda$s coming from the core, as compared to those coming from the corona, an amplification effect for the $\overline{\Lambda}$ with respect to that of $\Lambda$ polarization can occur. This amplification becomes more important for lower collision energies and quantitatively accounts for the $\Lambda$ and $\overline{\Lambda}$ polarizations measured by the STAR beam energy scan. 
\end{abstract}
\maketitle

The polarization asymmetry of a given baryon species produced in high-energy reactions is
defined as the ratio of the difference between the number of baryons with their spin pointing along and opposite to a given direction, to their sum.
This direction is usually chosen as either the baryon momentum or the normal to the production plane. In the former case one speaks of the longitudinal, whereas the latter is referred to as the transverse polarization. 

Among the baryons whose polarization properties can be studied, $\Lambda$ plays an important role. In addition of being the lightest hyperon with strange quark content, it has a self-analyzing polarization power due to its parity-violating weak decay, since the decay protons  follow preferentially the spin direction of the original $\Lambda$.
\begin{figure}[th!]
{\centering
\includegraphics[width=0.4\textwidth]{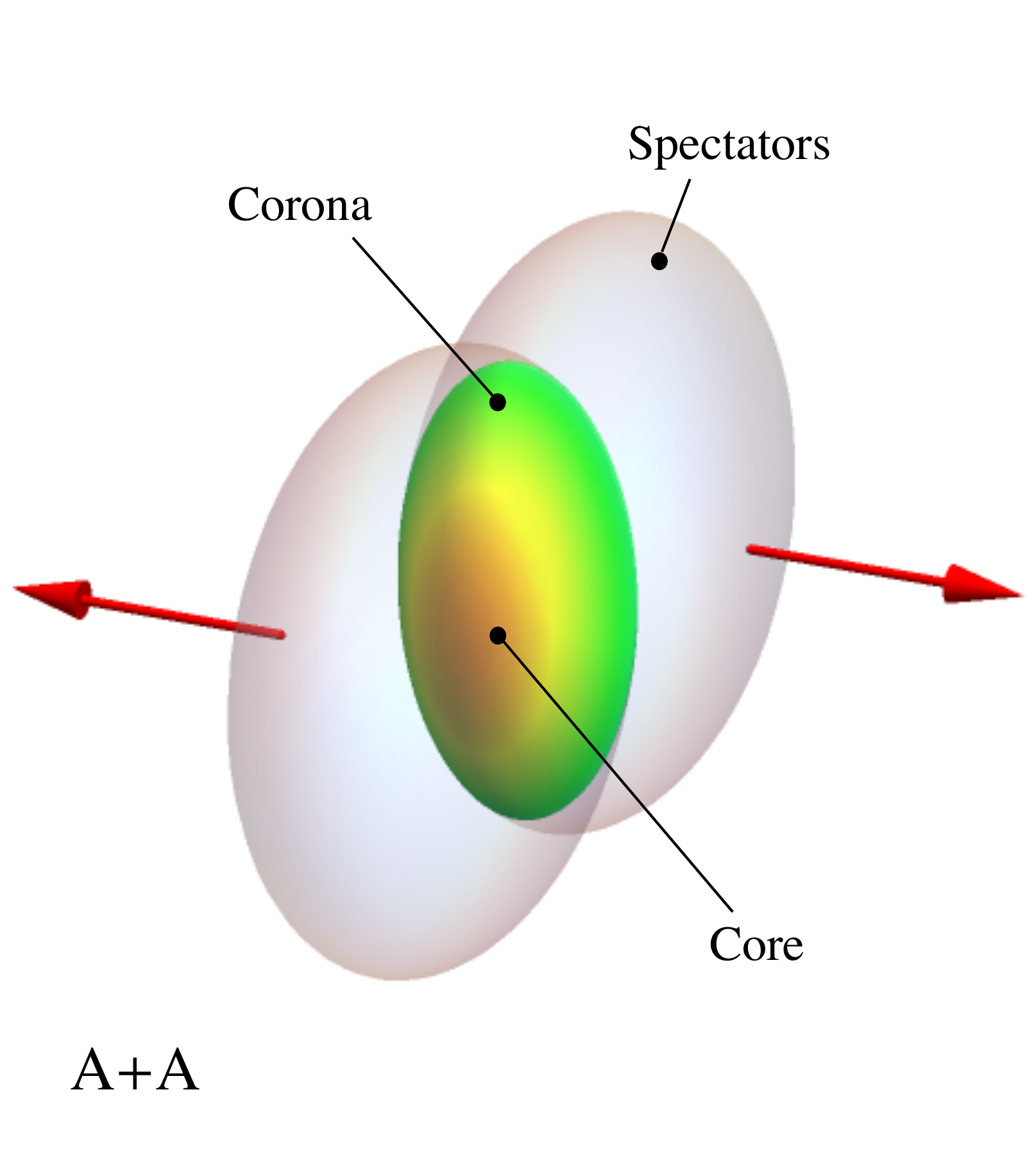}}
\caption{Illustration of a non-central heavy-ion collision of a symmetric system with impact parameter $b$. The core produces $\Lambda$s and $\overline{\Lambda}$s from QGP processes, whereas from the corona these particles are produced by n + n reactions.}
\label{fig1}
\end{figure}

$\Lambda$ and $\overline{\Lambda}$ polarization have been extensively studied, both from the experimental and the theoretical points of view. On the experimental side, these studies date back to the pioneering Fermilab measurements~\cite{Bunce}. $\Lambda$ and $\overline{\Lambda}$ appear polarized over a wide range of  collision energies and systems from p + p~\cite{Aad,Abelev,Felix,Felix2,Felix3,Heller,Erhan,Smith,Ramberg}, p + A~\cite{Agakishiev,Abt}, deep inelastic scattering~\cite{Airapetian,Astier,Adams} and even in $e^+ + e^-$~\cite{Buskulic,Guan} collisions.
The polarization mechanism is not well understood. A wealth of theoretical explanations have been put forward with varying degrees of success depending on the kind of colliding systems and energy ranges~\cite{Groom,Yen,Andersson,DeGrand,Panagiotou,Soffer,Gustafson,Ellis,Ellis2,Jaffe,Kotzinian,Florian,Florian2,Boros,Anselmino,Anselmino2,Ma,Alikhanov,Yang,Sun,Han}. 
In recent years, the interest on $\Lambda$ and $\overline{\Lambda}$ polarization has been further increased due to the possibility to link this observable to the properties of the medium produced in relativistic heavy-ion collisions~\cite{Jacob,Barros,Ladygin,Becattini1,Xie,Karpenko,Xie2,Liao,Liao2,Li,Karpenko2,Xia,Suvarieva}. For non-central reactions, the inhomogeneity of the matter density profile in the transverse plane produces the colliding region to develop an orbital angular momentum~\cite{Becattini2008}, quantified in terms of the {\it thermal vorticity}~\cite{Becattini2017}, defined as $
   \overline{\omega}_{\mu\nu}=\left(\partial_\nu\beta_\mu - \partial_\mu\beta_\nu\right)/2$,
where $\beta_\mu=u_\mu(x)/T(x)$, $u_\mu(x)$ is the local fluid four-velocity and $T(x)$ is the local temperature. In the non-relativistic limit and assuming global equilibrium, the thermal vorticity can be written as the ratio of a constant angular velocity and a constant temperature. By choosing the direction of reference as the angular momentum, which coincides with the normal to the reaction plane, it is possible to measure the so called {\it global polarization}.

The ALICE~\cite{ALICE} and STAR~\cite{STAR-Nature,STAR2} collaborations have reported results for global $\Lambda$ and $\overline{\Lambda}$ polarization. In particular, the STAR Beam Energy Scan (BES) has shown that as the collision energy decreases, the $\overline{\Lambda}$ polarization increases more steeply than the $\Lambda$ polarization. To explain this behavior, different space-time distributions and freeze-out conditions for $\Lambda$ and $\overline{\Lambda}$ have been invoked~\cite{Vitiuk}. Also, since the $s$ ($\bar{s}$)-quark, thought to be the main responsible for the $\Lambda$ ($\overline{\Lambda}$) polarization, has negative (positive) electric charge, it has been suggested that differences between the $\Lambda$ and $\overline{\Lambda}$ global polarization may be due to the strong, albeit short-lived, magnetic field that is produced in non-central collisions~\cite{Hai-Bo,Liao3,Liao4}. The possibility that $\Lambda$ and $\overline{\Lambda}$ align their spins with the direction of the angular momentum, during the life-time of the system created in the reaction, has been recently put on firmer grounds in Refs.~\cite{Ayala,newtau}.

In non-central collisions $\Lambda$ and $\overline{\Lambda}$ hyperons can be produced from different density zones within the interaction region. This scenario was put forward in Ref.~\cite{Ayala2}.
The $\Lambda$ and $\overline{\Lambda}$ polarization properties can therefore differ depending on whether these particles come from the denser (core) or less dense (corona) regions. In this work, we explore such two-component scenario. We show that since the ratio of the number of $\overline{\Lambda}$s to $\Lambda$s coming from the corona is less than 1, the global $\overline{\Lambda}$ polarization can be larger than the global $\Lambda$ polarization, in spite of the {\it intrinsic}, thermal vorticity-produced, $\Lambda$ polarization in the core being larger than the $\overline{\Lambda}$ polarization. This amplifying effect is favored when the number of $\Lambda$s coming from the core is smaller than the number of $\Lambda$s coming from the corona. The latter can happen for collisions with intermediate to large impact parameters, which at the same time, correspond to the kind of collisions that favor the development of a larger thermal vorticity.

Consider the scenario where in a peripheral heavy-ion reaction, the number of $\Lambda$s, come from two regions:  a high-density core and a less dense corona, such that $N_\Lambda=\NQGP + \NREC$, where $\NQGP$ is the number of produced $\Lambda$s coming from the core and $\NREC$ is the number of produced $\Lambda$s coming from the corona. These zones are illustrated in Fig.~\ref{fig1}. The subscripts ``QGP" and ``REC" refer to the kind of processes responsible for the production of these hyperons, that is, QGP in the core and recombination induced process in the corona, respectively. The subscript \lq\lq REC” refers to the name given in the early literature to the process whereby $\Lambda$s are produced in p + p collisions from reactions where a $u$-$d$ diquark picks up an $s$ quark from the sea, thus {\it recombining} to produce a $\Lambda$. These processes are thought to also produce polarization in these reactions. The expression for the $\Lambda$ and $\overline{\Lambda}$  polarization is given by
\begin{align}
\PLambda&=\frac{\
(\NupQGP+\NupREC)-(\NdownQGP+\NdownREC)}{\ 
(\NupQGP+\NupREC)+(\NdownQGP+\NdownREC)},\nonumber \\
\PLambdabar&=\frac{\ 
(\NbarupQGP+\NbarupREC)-(\NbardownQGP+\NbardownREC)}{\ 
(\NbarupQGP+\NbarupREC)+(\NbardownQGP+\NbardownREC)}.
\label{polLambda}
\end{align}
After a bit of straightforward algebra, we can express the $\Lambda$ and $\overline{\Lambda}$ polarization, Eq.~(\ref{polLambda}) as
\begin{align}
\PLambda &=
\frac{\left( \PLambdaREC +
\frac{\NupQGP - \NdownQGP}{\NREC}\right)}{ \left( 1 +
\frac{\NQGP}{\NREC}\right)
}, \nonumber\\
\PLambdabar&=\frac{\left( \PLambdabarREC + 
\frac{\NbarupQGP - \NbardownQGP}{\NREC}\right)}{ \left( 1 + 
\frac{\NbarQGP}{\NbarREC}\right)
},
\end{align}
where 
\begin{eqnarray}
   \!\!\!\!\!\PLambdaREC=\frac{\NupREC - \NdownREC}{\NupREC + \NdownREC},\
   \PLambdabarREC=\frac{\NbarupREC - \NbardownREC}{\NbarupREC + \NbardownREC}.
\label{PREC}
\end{eqnarray}
Notice that $\PLambdaREC$ and $\PLambdabarREC$ refer to the polarization along the global angular momentum produced in the corona. Although nucleons colliding in this region partake of the vortical motion, reactions in cold nuclear matter are less efficient to align the spin in the direction of the angular momentum than in the QGP. Thus, as a working approximation we set $\PLambdaREC=\PLambdabarREC=0$ to write
\begin{eqnarray}
\!\!\!\!\!\PLambda=\frac{\left( \frac{\NupQGP - \NdownQGP}{\NREC}\right)}{ \left( 1 + \frac{\NQGP}{\NREC}\right)},\
\PLambdabar=\frac{\left(
\frac{\NbarupQGP - \NbardownQGP}{\NbarREC}\right)}{ \left( 1 + 
\frac{\NbarQGP}{\NbarREC}\right)}.
\label{polsPREC=0}
\end{eqnarray}
\begin{figure}[t!]
{\centering
\includegraphics[width=0.47\textwidth]{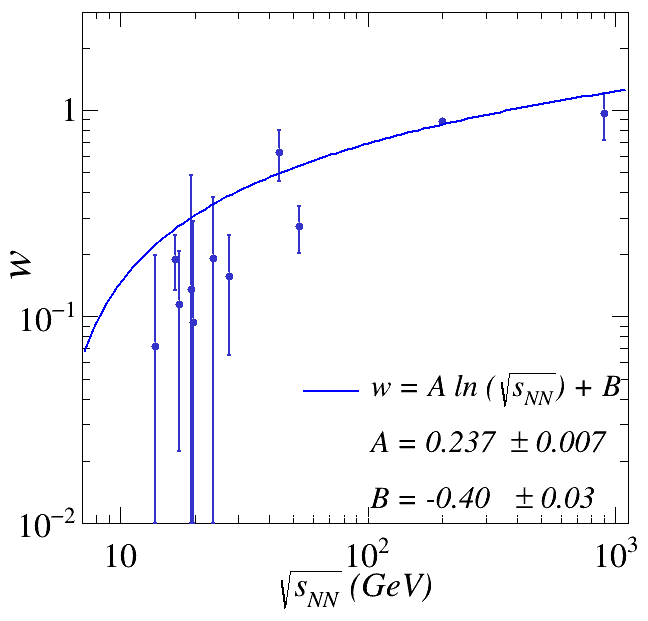}}
\caption{Experimental data on the ratio $w=\NbarREC/\NREC$ obtained from p + p collisions at different energies~\cite{Gazdzicki:1996pk,Chapman:1973fn,Brick:1980vj,Hohne:2003bca,Baechler:1991pp,Charlton:1973kw, Lopinto:1980ct, Kichimi:1979te, Busser:1975tj, Erhan:1979ba, Abelev:2006cs, Abbas:2013rua}. Shown is also the fit describing the data. Notice that the ratio is smaller than 1 except for the largest collision energy considered.}
\label{fig2}
\end{figure}
However, since reactions in the core are more efficient to align particle spin to global angular momentum, one expects that the {\it intrinsic} global $\Lambda$ and $\overline{\Lambda}$ polarizations namely, 
\begin{eqnarray}
    z&=& \frac{(\NupQGP- \NdownQGP)}{\NQGP} \nonumber\\
   \bar{z}&=& \frac{(\NbarupQGP - \NbardownQGP)}{ \NbarQGP}\simeq \frac{(\NbarupQGP - \NbardownQGP)}{\NQGP},
\label{aprox2}
\end{eqnarray}
are finite, albeit small. In Eq.~(\ref{aprox2}), we have used that in the QGP one expects $\NbarQGP \simeq \NQGP$. Therefore, Eq.~(\ref{polsPREC=0}) can be written as
\begin{eqnarray}
\mathcal{P}^\Lambda=\frac{z\frac{
N_{\Lambda\ {\mbox{\tiny{QGP}}}} }{N_{\Lambda\ {\mbox{\tiny{REC}}}}}}{ \left( 1 + \frac{N_{\Lambda\ {\mbox{\tiny{QGP}}}}}{N_{\Lambda\ {\mbox{\tiny{REC}}}}}\right)},\ \ \ \
\mathcal{P}^{\overline{\Lambda}}=\frac{\bar{z}
\frac{N_{\Lambda\ {\mbox{\tiny{QGP}}}} }{N_{\overline{\Lambda}\ {\mbox{\tiny{REC}}}}}}{ \left( 1 + 
\frac{N_{\Lambda\ {\mbox{\tiny{QGP}}}}}{N_{\overline{\Lambda}\ {\mbox{\tiny{REC}}}}}\right)}.
\label{withz}
\end{eqnarray}

Cold nuclear matter collisions in the corona are expected to produce more $\Lambda$s than $\overline{\Lambda}$s, since these processes are related to p + p reactions, where three anti-quarks coming from the sea are more difficult to produce than only one $s$. Then, we can write
\begin{figure}[bh!]
{\centering
\includegraphics[width=0.45\textwidth]{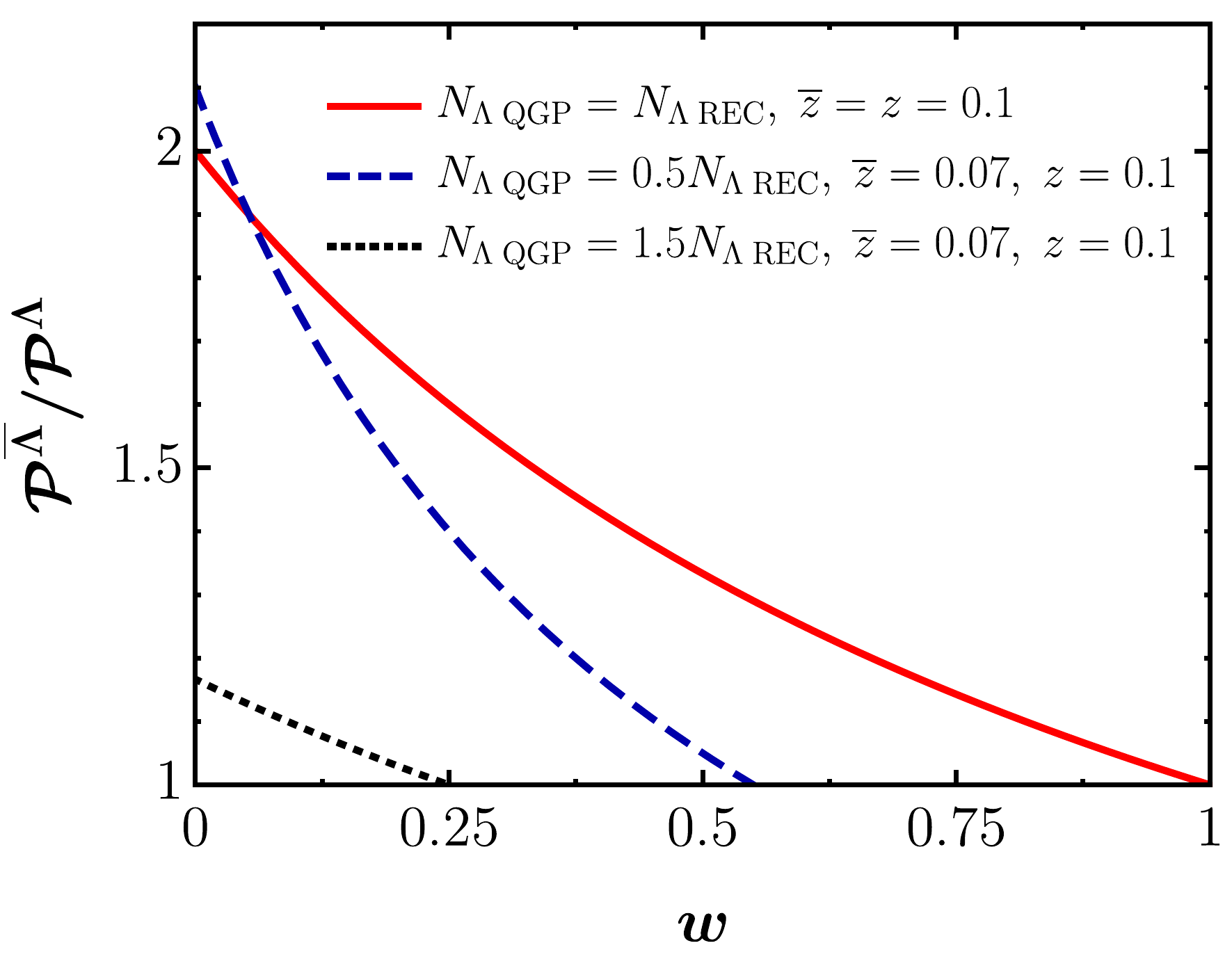}}
\caption{$\PLambdabar/\PLambda$ as a function of $w$. A range of $w$ values for which $\PLambdabar/\PLambda>1$ exists whose width depends on the relation between $\NQGP$, $\NREC$, $\bar{z}$ and $z$.}
\label{fig3}
\end{figure}
$\NbarREC \equiv w\NREC$. Notice that in contrast, given the assumption that $ N_{\overline{\Lambda}\ {\mbox{\tiny{QGP}}}}\simeq N_{\Lambda\ {\mbox{\tiny{QGP}}}}$, no such suppression factor similar to $w$ needs to be introduced in the QGP (core) region. Thus
\begin{eqnarray}
\mathcal{P}^\Lambda=\frac{z\frac{
N_{\Lambda\ {\mbox{\tiny{QGP}}}} }{N_{\Lambda\ {\mbox{\tiny{REC}}}}}}{ \left( 1 + \frac{N_{\Lambda\ {\mbox{\tiny{QGP}}}}}{N_{\Lambda\ {\mbox{\tiny{REC}}}}}\right)},\ \ \
\mathcal{P}^{\overline{\Lambda}}=\frac{\left(\frac{\bar{z}}{w}\right)
\frac{N_{\Lambda\ {\mbox{\tiny{QGP}}}} }{N_{\Lambda\ {\mbox{\tiny{REC}}}}}}{ \left( 1 + 
\left(\frac{1}{w}\right)
\frac{N_{\Lambda\ {\mbox{\tiny{QGP}}}}}{N_{\Lambda\ {\mbox{\tiny{REC}}}}}\right)},
\label{withw}
\end{eqnarray}
where the energy dependent coefficient $w$ is expected to be also smaller than 1. This expectation is in fact met, as shown in Fig.~\ref{fig2}, where $w$ is plotted as a function of $\sqrt{s_{NN}}$, using data compiled from different experimental results for p + p collisions~\cite{Gazdzicki:1996pk,Chapman:1973fn,Brick:1980vj,Hohne:2003bca,Baechler:1991pp,Charlton:1973kw, Lopinto:1980ct, Kichimi:1979te, Busser:1975tj, Erhan:1979ba, Abelev:2006cs, Abbas:2013rua}. 
The function describing the experimental points corresponds to a fit given by
$w(\sqrt{s_{NN}})=A\ln{\sqrt{s_{NN}}}+B$, with $A=0.237 \pm 0.007$ and $B = -0.40 \pm 0.03$. Notice that the $\Lambda$ and $\overline{\Lambda}$ polarization depend, in addition of $z$, $\bar{z}$ and $w$, also of the ratio $\NQGP/\NREC$. Although $\bar{z}$ is expected to be smaller than $z$, the amplifying effect from the factor $1/w>1$ produces that $\PLambdabar>\PLambda$ for a range of $w$ values. This is illustrated in Fig.~\ref{fig3} where we plot  $\PLambdabar/\PLambda$ as a function of $w$. In the extreme situation where $\bar{z}=z$ and $\NQGP/\NREC=1$, $\PLambdabar/\PLambda$ is always larger than 1 for $0<w<1$. For a more realistic scenario with $\bar{z}<z$ and with $\NQGP/\NREC$ smaller than 1, there is still a range of $w$ values for which $\PLambdabar/\PLambda$ is larger than 1. This region shrinks when $\NQGP/\NREC>1$.
\begin{figure}[t!]
{\centering
\includegraphics[width=0.45\textwidth]{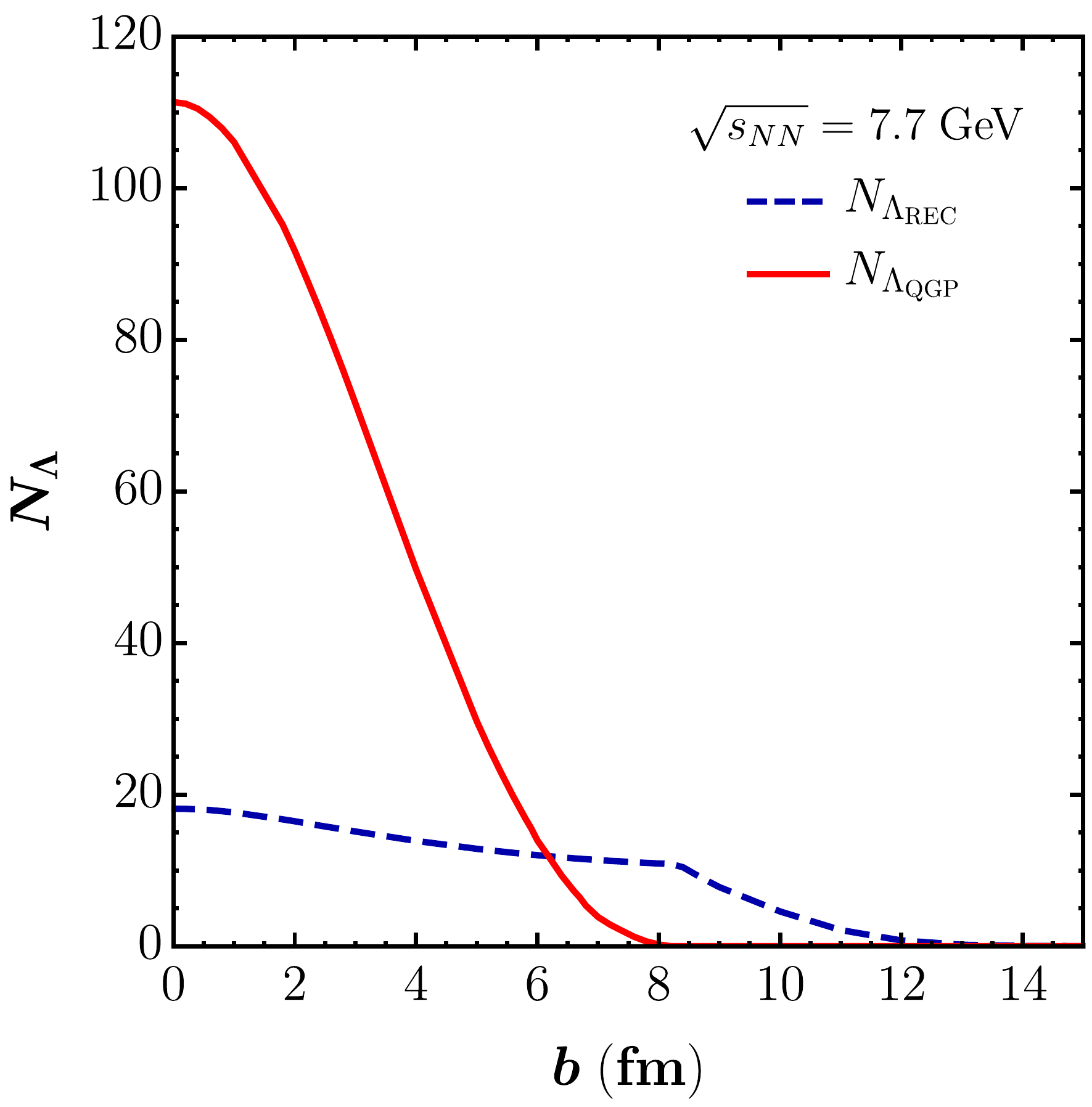}}
\caption{Example of the number of $\Lambda$s originating from the core $\NQGP$ and the corona $\NREC$ as a function of impact parameter $b$ for $\sqrt{s_{NN}}=7.7$ GeV. Notice that for $b\gtrsim 6$ fm, $\NQGP/\NREC<1$.}
\label{fig4}
\end{figure}

In order to check whether non-central collisions at different energies and impact parameters favor a scenario where $\NQGP/\NREC \lesssim 1$ and thus $\PLambdabar>\PLambda$, we proceed to study $\Lambda$ production in the QGP and REC regions. Recall that the average number of strange quarks produced in the QGP scales with the number of participants $\NpQGP$ in the
collision roughly as
$
\langle s\rangle = \NQGP = c\, \NpQGP^2,
$
where in Ref.~\cite{Letessier}, $c$ is found to be in the range $0.001\leq c\leq 0.005$, assuming, for the sake of simplicity, that as a result of hadronization, only $\Lambda$s and $\overline{\Lambda}$s are obtained from these produced
$s$-quarks. Hereafter we work explicitly with a proportionality factor $c=0.0025$ corresponding to an intermediate value of the above range to account for the fact that $\Lambda$s are not the only strange hadrons produced in the reaction. The number of $\Lambda$s originating in the QGP can be computed from the relation
\begin{eqnarray}
   \NpQGP = \int d^2s\ n_\text{p}(\Vec{s},\Vec{b})\,\theta \left[n_\text{p}(\Vec{s},\Vec{b})-n_c\right],
\label{numpart}
\end{eqnarray}
where the density of participants $n_\text{p}$ is given in terms of the thickness functions $T_A$ and $T_B$ of the colliding system A+B as
\begin{eqnarray}
   n_\text{p}(\Vec{s},\Vec{b})&=&T_A(\vec{s}\,)[1-e^{-\sigma_{NN}(\sqrt{s_{NN}})T_B(\vec{s}-\vec{b})}]\nonumber\\
   &+&T_B(\vec{s}-\vec{b})[1-e^{-\sigma_{NN}(\sqrt{s_{NN}})T_A(\vec{s})}],
\label{np}
\end{eqnarray}
\begin{figure}[t!]
{\centering
\includegraphics[width=0.47\textwidth]{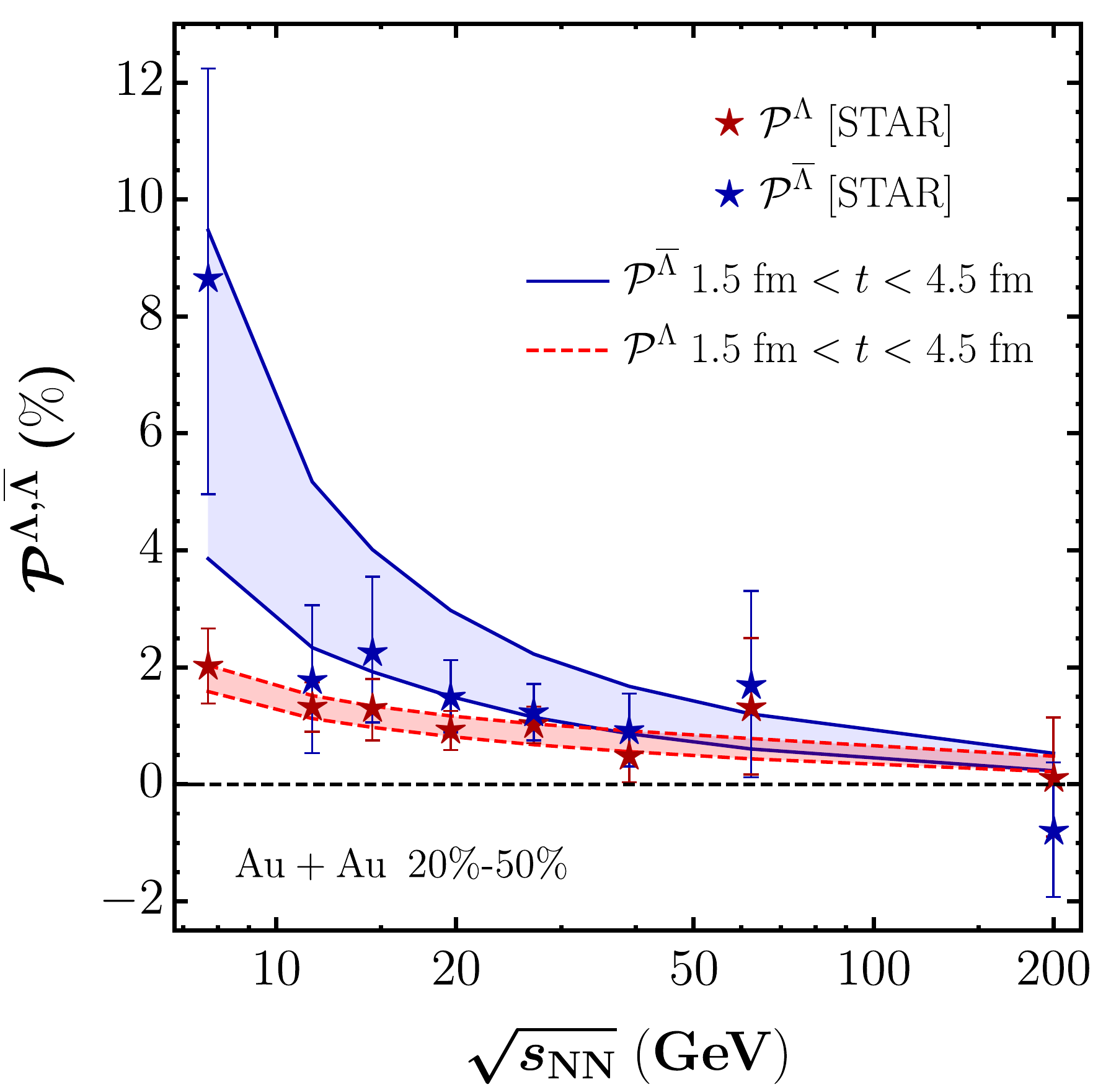}}
\caption{Computed $\Lambda$ and $\overline{\Lambda}$ polarization compared to data from the BES~\cite{STAR-Nature}. The bands correspond to the polarization obtained for 1.5 fm $<t<$ 4.5 fm.}
\label{fig5}
\end{figure}
with $\vec{b}$ the vector directed along the impact parameter on the nuclei overlap area and $\sigma_{NN}$ the collision energy-dependent nucleon-nucleon cross-section~\cite{PDG,ALICE:sigmaNN}. $n_c=3.3$ fm$^{-2}$ is the critical density of participants above which the QGP can be produced~\cite{Blaizot}. The thickness function $T_A$ is given by
\begin{eqnarray}
   T_A(\vec{s}\,)=\int_{-\infty}^{\infty}\rho_A(z,\vec{s}\,)\;dz,
\label{TA}
\end{eqnarray}
where we take as the nuclear density $\rho_A$ a Woods-Saxon profile
with a skin depth $a=0.41$ fm~\cite{SkinDepth} and a radius for a nucleus with mass number $A$ of $R_A=1.1A^{1/3}$ fm. On the other hand, the number of $\Lambda$s produced in the corona can be written as
\begin{align}
   \!\!\!\!\NREC &= \sigma_{NN}^\Lambda\left(\sqrt{s_{NN}}\right)
   \int d^2s\; T_B(\vec{b}-\vec{s}) \nonumber \\
   &\qquad \qquad \qquad \times T_A(\vec{s}\,)\,\theta \left[n_c-n_\text{p}(\Vec{s},\Vec{b})\right],
\label{partper}
\end{align}
where $\sigma_{NN}^\Lambda$ is the collision-energy dependent cross-section for $\Lambda$ production in p + p reactions. We obtain this function fitting experimental data from Refs.~\cite{Brick:1980vj,Kichimi:1979te,Erhan:1979ba,Jaeger:1974in,Blobel:1973jc,Drijard:1981wg}. The fit is given by
$\sigma_{NN}^\Lambda\left(\sqrt{s_{NN}}\right) = C\ln\sqrt{s_{NN}}+D$, with $C=1.67\pm0.05$ mb and $D=-1.60\pm0.08$ mb. Figure~\ref{fig4} shows an example of $\NQGP$ and $\NREC$ for a collision energy $\sqrt{s_{NN}}=7.7$ GeV as a function of impact parameter. Notice that in this case the ratio $\NQGP/\NREC$ becomes smaller than 1 for  $b\gtrsim 6$ fm. 

We now put together all these ingredients to study $\Lambda$ and $\overline{\Lambda}$ polarization as functions of the collision energy. We resort to the results of Ref.~\cite{newtau} where the  relaxation times $\tau$ and $\bar{\tau}$ for the alignment between the spin of a $s$ or a $\bar{s}$ with the thermal vorticity, respectively, are computed as functions of the collision energy. When the $s$ and $\bar{s}$ polarization translate into the $\Lambda$ and $\overline{\Lambda}$ polarization, respectively, during the hadronization process, the intrinsic polarization $z$ and $\bar{z}$ can be computed from these relaxation times as $z=1-\exp{(-t/\tau)}$ and $\bar{z}=1-\exp{(-t/\bar{\tau})}$, as functions of the $\Lambda$ and $\overline{\Lambda}$ formation time $t$ within the QGP.
Figure~\ref{fig5} shows the $\Lambda$ and $\overline{\Lambda}$ polarization thus computed compared to results from the BES~\cite{STAR-Nature}. The band shows the result of the calculation when the time for $\Lambda$ and $\overline{\Lambda}$ formation within the QGP is taken to lie between 1.5 fm $< t <$ 4.5 fm for $b=8$ fm, corresponding to the average impact parameter in the 20-50\% centrality range where data are taken. Notice that the polarization data is well described by the calculation over the entire collision energy range. 

In conclusion, we have shown that when the source of $\Lambda$s and $\overline{\Lambda}$s in a semi-central heavy-ion collision is modeled as composed of a high-density core and a less dense corona, their global  polarization properties as a function of the collision energy can be understood. It is in the core, that the QGP is produced and thus the requirement that the density of participants is higher than a critical value $n_c$. However, once the system evolves, this region becomes the one with low baryon density. The overall high baryon density comes from the corona, which at lower energies corresponds to a larger volume than at higher energies. In the QGP it is equally as easy to produce $\Lambda$s as it is to produce $\overline{\Lambda}$s, given that in this region quarks and antiquarks are thought to be {\it freely} roaming around and three antiquarks ($\bar{u}$, $\bar{d}$, $\bar{s}$) can find each other as easily as three quarks ($u,\ d,\ s$) to form $\overline{\Lambda}$s and $\Lambda$s, respectively. On the other hand in the corona (high baryon density), reactions are similar to those in p + p collisions, where we know that it is easier to produce $\Lambda$s than $\overline{\Lambda}$s. This last fact is supported by our compilation of experimental results for the ratio $w$ which satisfies $w<1$. 

When a larger abundance of $\Lambda$s as compared to $\overline{\Lambda}$s in the corona is combined with a smaller number of $\Lambda$s coming from the core as compared to those coming from the corona -- which happens for semi-central to peripheral collisions -- an amplification effect for the $\overline{\Lambda}$ polarization can occur, in spite of the {\it intrinsic} $\Lambda$ polarization $z$ being larger than the {\it intrinsic} $\overline{\Lambda}$ polarization $\bar{z}$. This amplification is more prominent for lower collision energies. A more detailed analysis to relax the approximation of equal number of $\Lambda$s and $\overline{\Lambda}$s produced in the QGP, and a vanishing polarization in the corona, as well as including a weighted average over contributing impact parameters
and formation times, is currently being performed and will be reported elsewhere. 

\section*{Acknowledgments}

A.A. thanks F. Wang for helpful comments. Support for this work has been received in part by UNAM-DGAPA-PAPIIT grant number IG100219 and by Consejo Nacional de Ciencia y Tecnolog\'ia grant numbers A1-S-7655 and A1‐S‐16215. I.M. acknowledges support from a postdoctoral fellowship granted by Consejo Nacional de Ciencia y Tecnolog\'ia.

\end{document}